%% file: ms.tex
\documentclass[aip, a4paper, amsfonts, amssymb, amsmath, reprint, showkeys, nofootinbib, twoside]{revtex4-2}
\usepackage[english]{babel}
\usepackage[utf8]{inputenc}
\usepackage[colorinlistoftodos, color=green!40, prependcaption]{todonotes}

\input{preamble.tex}
\usepackage[pdftex, pdftitle={Article}, pdfauthor={Author}]{hyperref} 

\begin{document}

\title{Ternary Metal Oxide Substrates for Superconducting Circuits}

\author{Zach Degnan}
    \affiliation{School of Mathematics and Physics,The University of Queensland, Brisbane, QLD 4072, Australia}

\author{Xin He}
    \affiliation{School of Mathematics and Physics,The University of Queensland, Brisbane, QLD 4072, Australia}
    \affiliation{ARC Centre of Excellence for Engineered Quantum Systems, St Lucia, Queensland 4072, Australia}

\author{Alejandro Gomez Frieiro}
    \affiliation{School of Mathematics and Physics,The University of Queensland, Brisbane, QLD 4072, Australia}
    \affiliation{ARC Centre of Excellence for Engineered Quantum Systems, St Lucia, Queensland 4072, Australia}

\author{Yauhen P. Sachkou}
    \affiliation{School of Mathematics and Physics,The University of Queensland, Brisbane, QLD 4072, Australia}
    \affiliation{ARC Centre of Excellence for Engineered Quantum Systems, St Lucia, Queensland 4072, Australia}

\author{Arkady Fedorov}
    \email[Correspondence email address: ]{a.fedorov@uq.edu.au}
    \affiliation{School of Mathematics and Physics,The University of Queensland, Brisbane, QLD 4072, Australia}
    \affiliation{ARC Centre of Excellence for Engineered Quantum Systems, St Lucia, Queensland 4072, Australia}   

\author{Peter Jacobson}
    \email[Correspondence email address: ]{p.jacobson@uq.edu.au}
    \affiliation{School of Mathematics and Physics,The University of Queensland, Brisbane, QLD 4072, Australia}

\date{\today} 

\begin{abstract}
\input{abstract.tex}
\end{abstract}


\maketitle

\input{intro.tex}
\input{experiment.tex}
\input{results.tex}

\input{discussion.tex}
\input{conclusion.tex}
\input{acknowledgements.tex}

\bibliography{ms}

\begin{figure}
    \includegraphics[width=\linewidth]{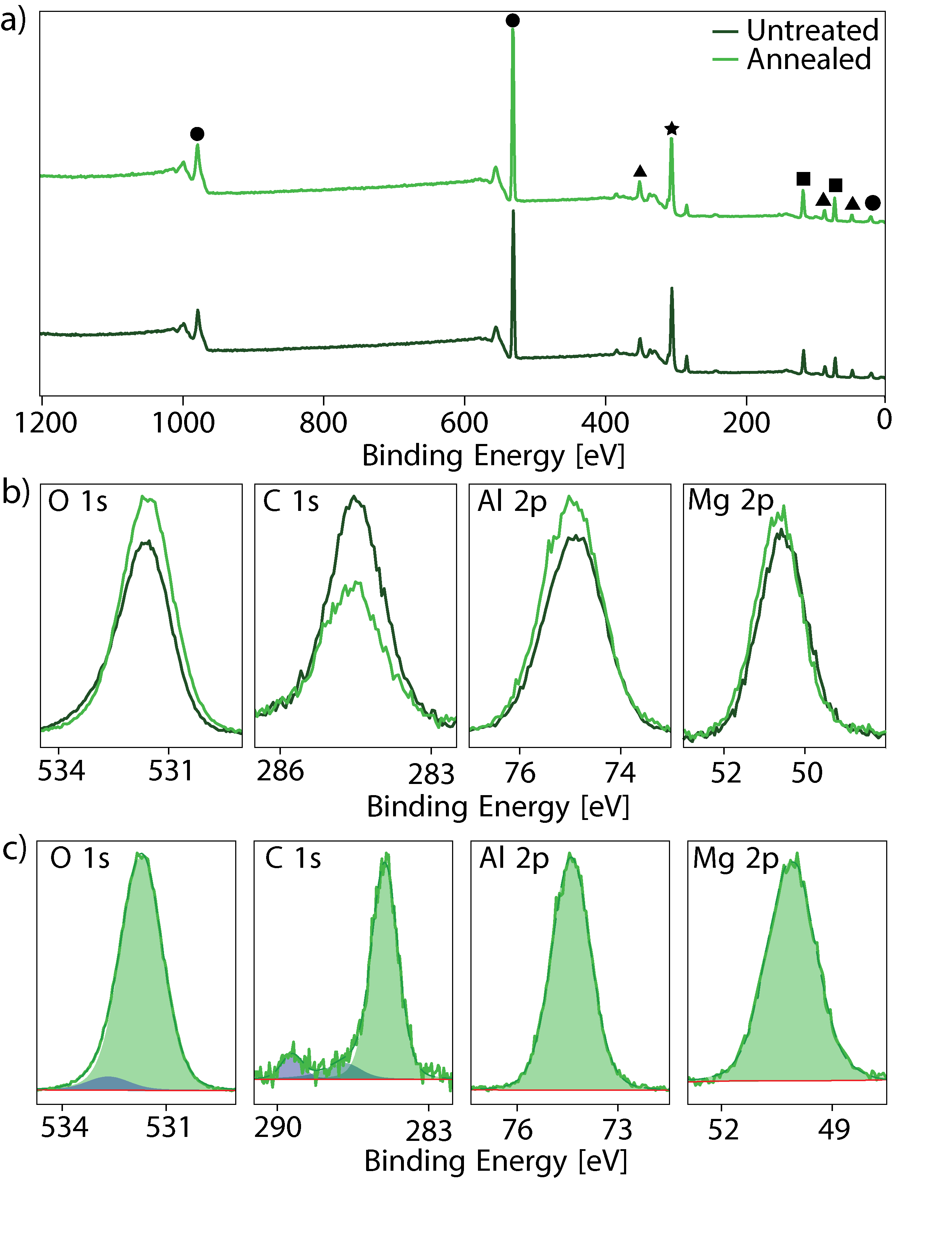}
    \caption{XPS results for untreated (dark green) and tube-furnace annealed (light green) MgAl$_2$O$_4$(100) substrates. \textbf{a)} Survey scans of untreated and annealed substrates. The oxygen (circles), aluminum (squares), magnesium (triangles) and carbon (stars) peaks are labelled. \textbf{b)} High-resolution scans of characteristic spinel and carbon XPS peaks before and after annealing. \textbf{c)}  Components of characteristic MgAl$_2$O$_4$ peaks on an annealed sample prior to aluminum deposition. The main O 1s component corresponds to the lattice oxygen, the component at higher binding energy is attributed to surface hydroxyls. The C 1s spectrum corresponds to predominantly sp$^3$ carbon with minor C-O and C=O contributions. The Al 2p and Mg 2p peaks were fit with a single component.}
    \label{XPSMAOfig}
\end{figure}

\begin{figure}
    \includegraphics[width=\linewidth]{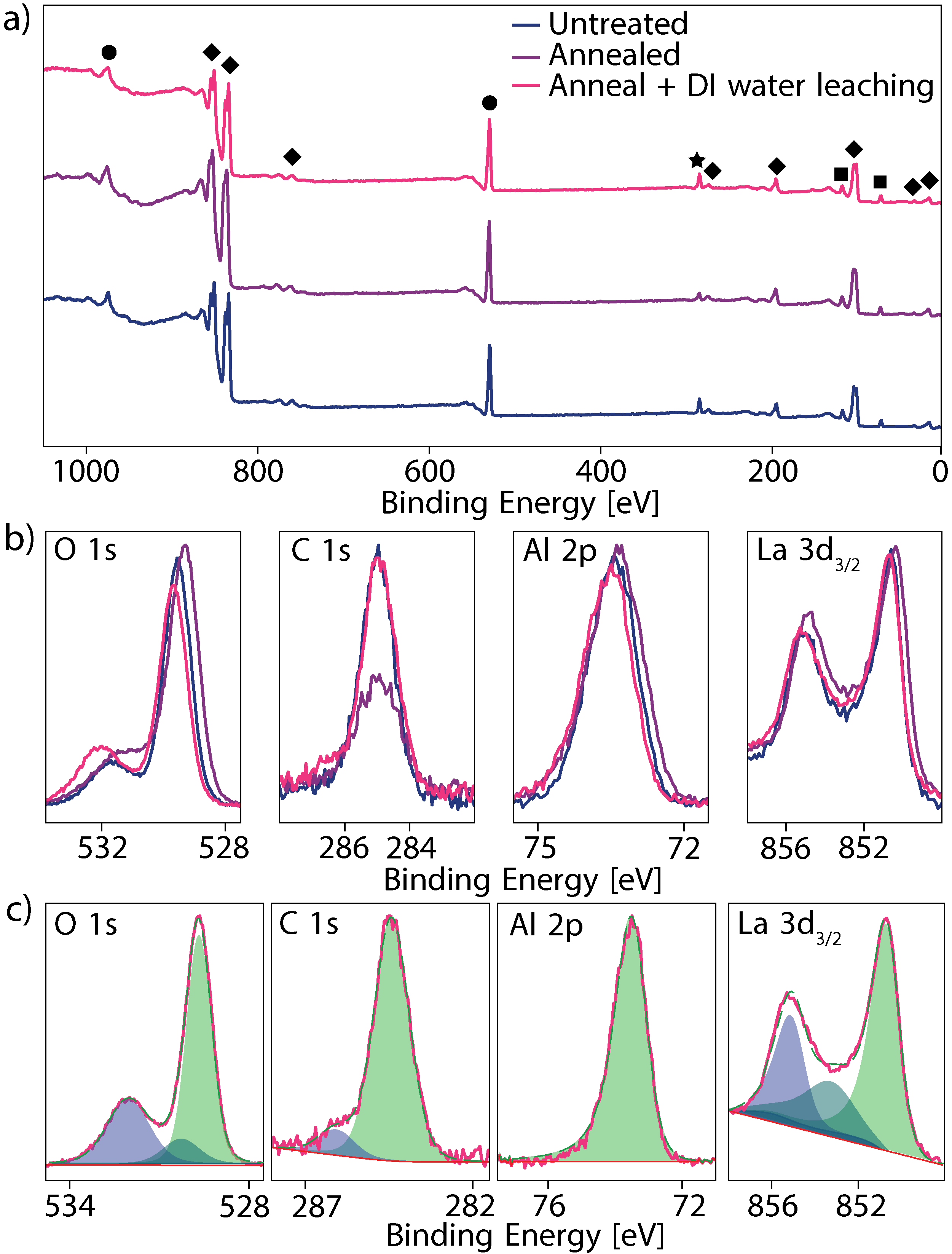}
    \caption{XPS results for LaAlO$_3$(100) substrates at three stages of preparation: untreated (blue), annealed at 1473 K (purple), and DI water leached for 120 minutes (pink). \textbf{a)} Survey scans at all three stages of preparation. The oxygen (circles), aluminum (squares), lanthanum (diamonds) and carbon (stars) peaks are labelled. \textbf{b)} Comparison of high-resolution scans of characteristic lanthanum aluminate and adventitious carbon XPS peaks. \textbf{c)}  High-resolution scan including components, component envelope (dotted green line) and background (red line)  of the fully treated (annealed and DI water leached) sample before aluminum deposition. The Al 2p peak was fit with a single component. The La 3d level is fit as described by Sunding et al. \cite{sunding2011}}
    \label{XPSLAOfig}
\end{figure}



\begin{figure}
    \includegraphics[width=\linewidth]{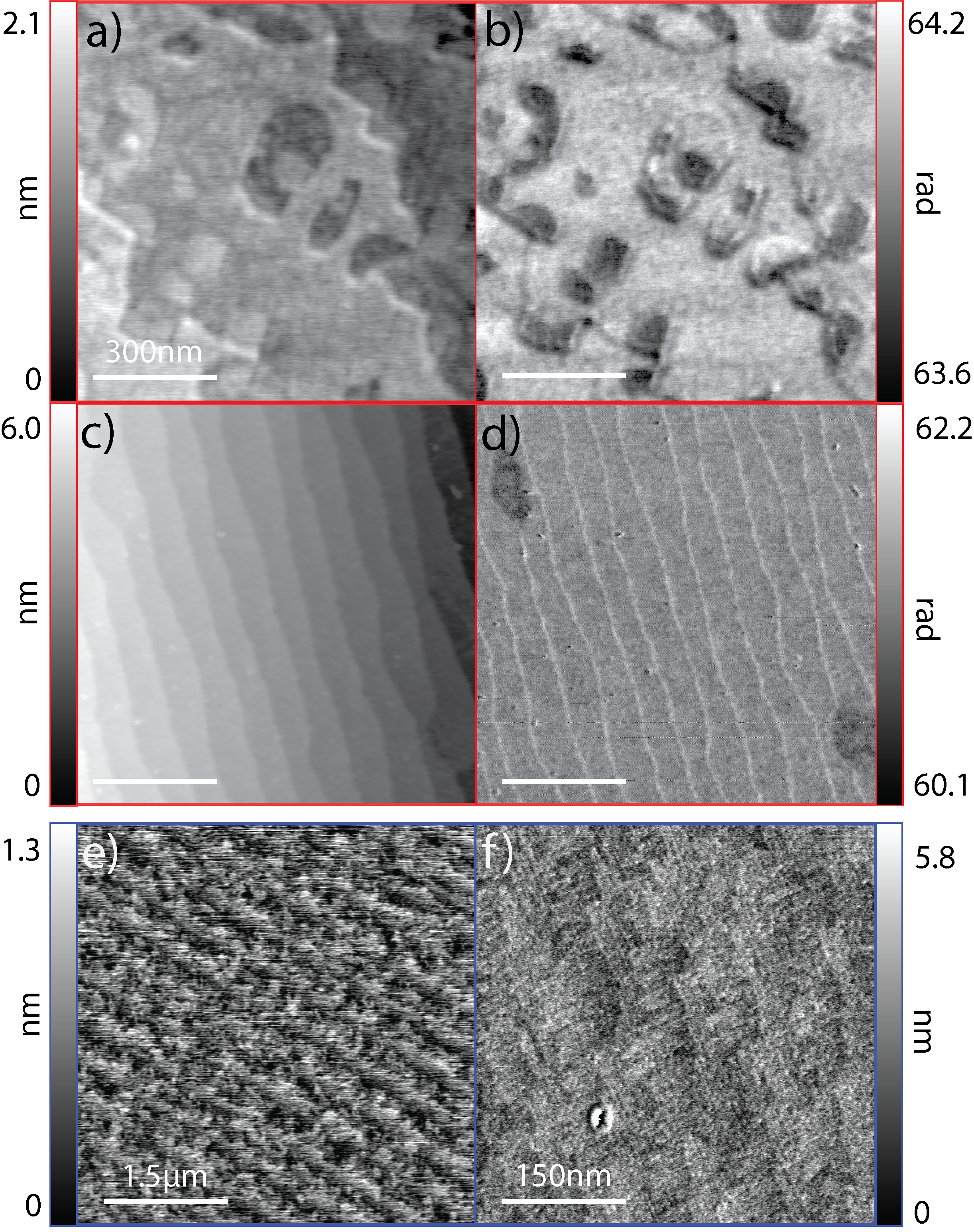}
    \caption{AFM height and phase images of LaAlO$_3$(100) (red) and MgAl$_2$O$_4$(100) (blue) substrates. \textbf{a)} Height and \textbf{b)} phase images of lanthanum aluminate after annealing at 1473 K, and \textbf{c-d)} after DI water leaching. All AFM images of LaAlO$_3$(100) are 1 $\times$ 1 $\mu$m$^2$. \textbf{e)} 2.5 $\times$ 2.5 $\mu$m$^2$ AFM image of MgAl$_2$O$_4$(100) showing a disordered step structure after annealing.  \textbf{f)} 500 $\times$ 500 nm$^2$ AFM image after annealing at 1373 K near a surface defect showing faint cross-hatching. Images of MgAl$_2$O$_4$(100) were high pass filtered.}
    \label{AFMfig}
\end{figure}

\begin{figure}
    \includegraphics[width=\linewidth]{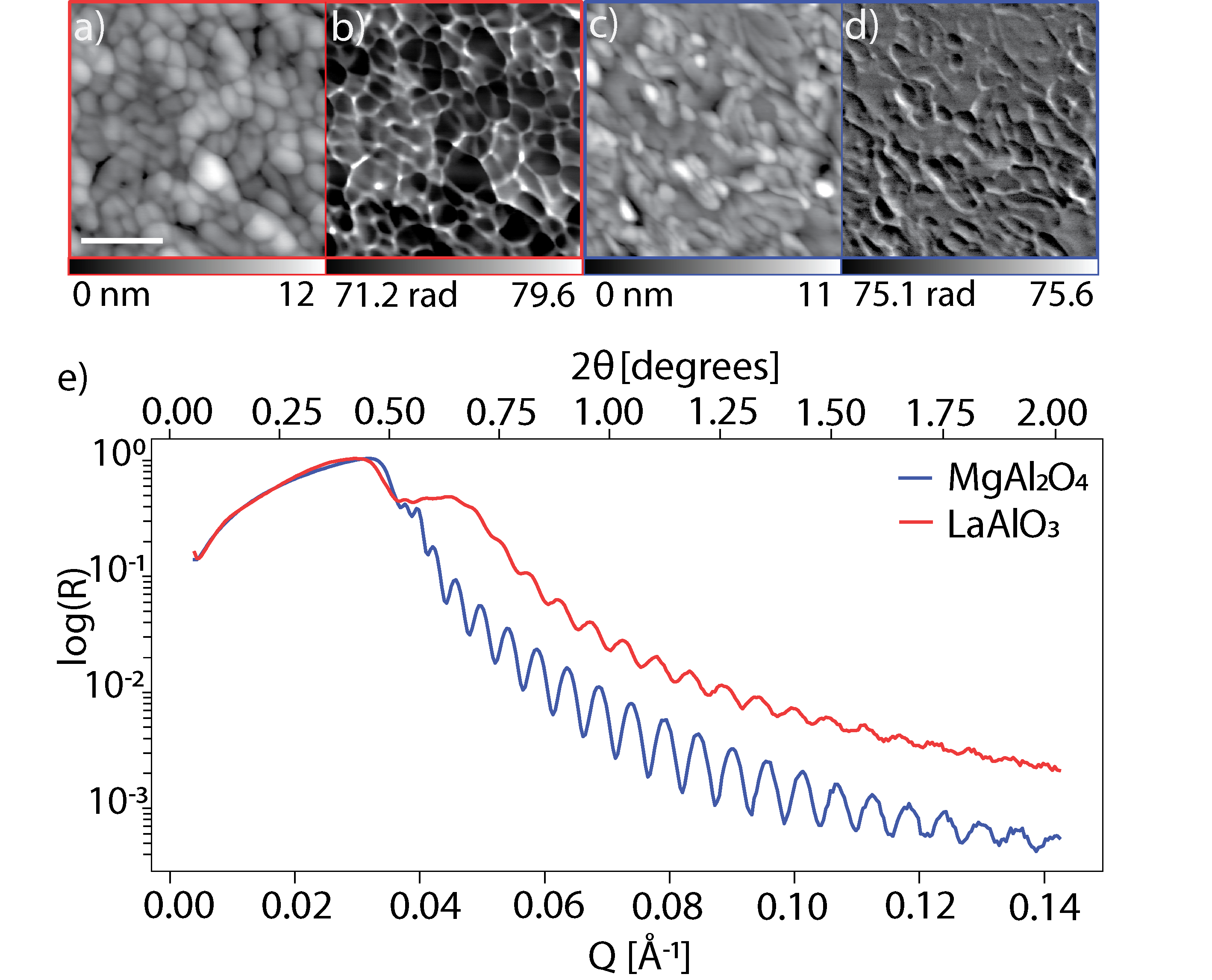}
    \caption{1 × 1 $\mu$m$^2$ AFM \textbf{a)} height and \textbf{b)} phase images of 100 nm thick Al films grown on LaAlO$_3$(100) (red) and \textbf{c-d)} MgAl$_2$O$_4$(100) (blue) substrates by electron beam deposition at a rate of 0.5 nm/s. The scale bar corresponds to 300 nm. \textbf{e)} XRR measurements of Al films on LaAlO$_3$(100) (red) and MgAl$_2$O$_4$(100) (blue) substrates.}
    \label{XRRfig}
\end{figure}

\begin{figure}
    \includegraphics[width=\linewidth]{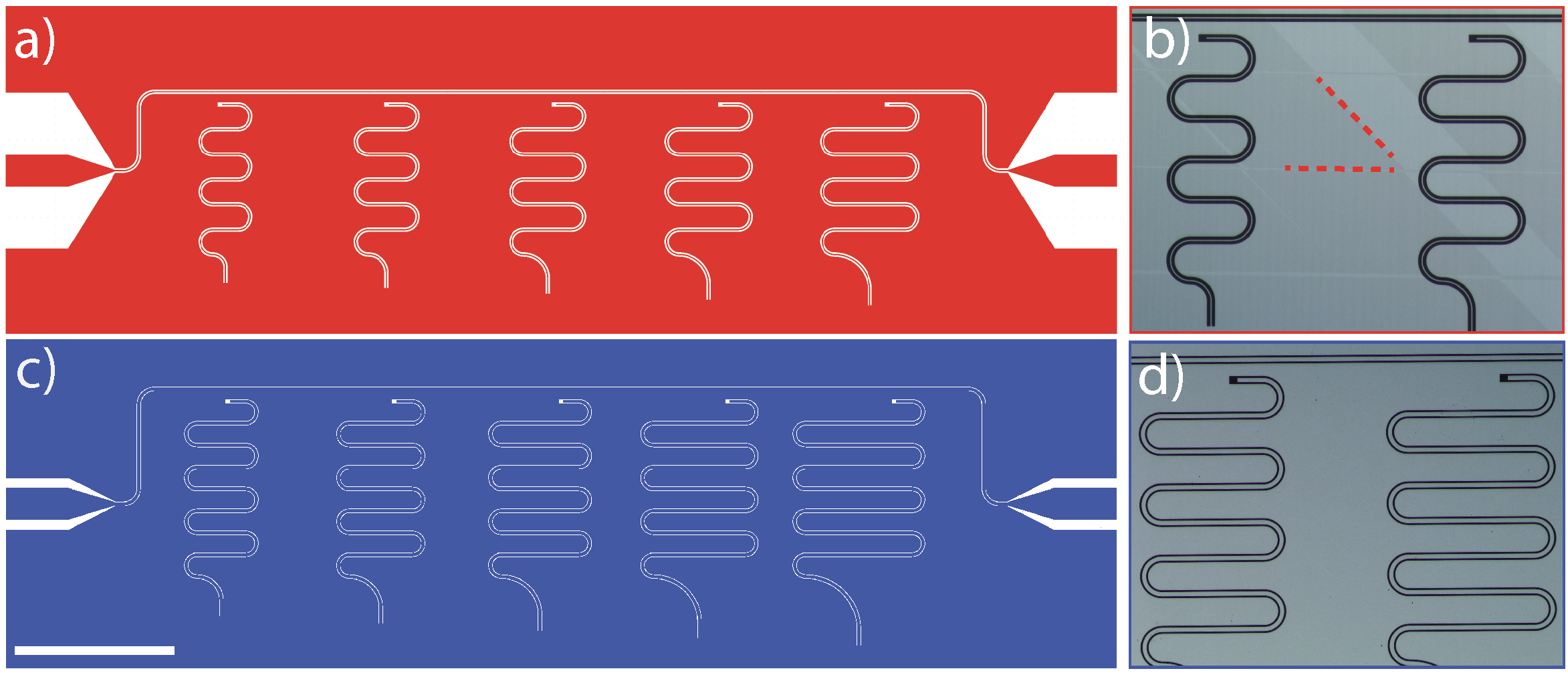}
    \caption{\textbf{a)} Digital render of the LaAlO$_3$ chip design and \textbf{b)} optical microscope image of the two left most resonators after Al etching. The dotted red lines show the visible twin boundaries in LaAlO$_3$. \textbf{c-d)} Digital render and optical microscope image of MgAl$_2$O$_4$ chips. The scale bar corresponds to 1 mm.}
    \label{geomfig}
\end{figure}

\begin{figure}
    \includegraphics[width=\linewidth]{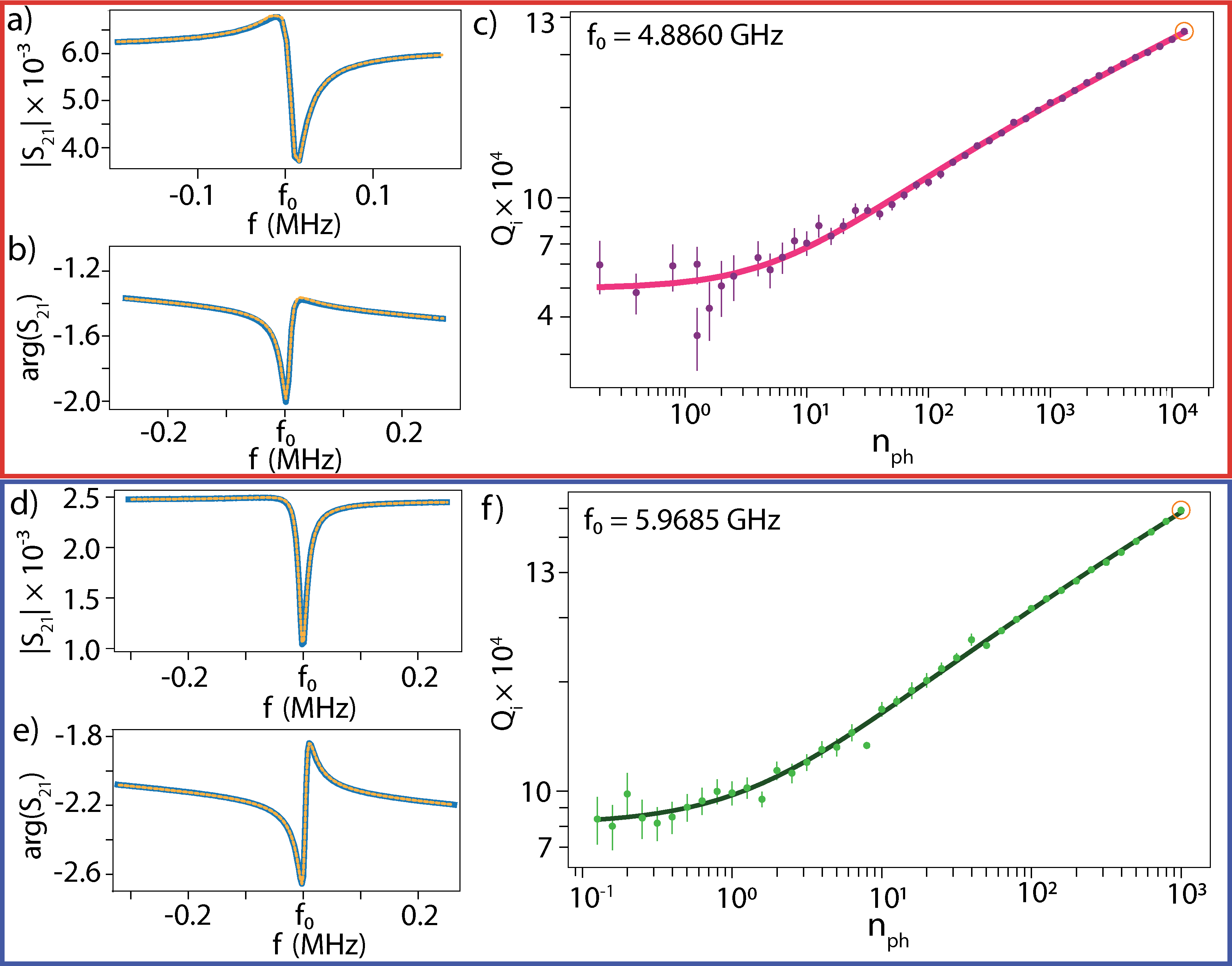}
    \caption{Low temperature measurements of devices across a range of input powers. \textbf{a-b)} The on-resonance amplitude and phase of the S$_{21}$ transmission through a resonator on an LaAlO$_3$(100) (red) substrate.  \textbf{c)} The power dependent internal Q-factor ($Q_i$) of the same resonator with a natural resonance frequency of $f_0 = 4.8860$ GHz. The resonance shown in a-b) corresponds to the circled data-point in c), with an input power of $\langle n_{ph} \rangle = 10^4$. \textbf{d-f)} Internal Q-factor measurements of an MgAl$_2$O$_4$ (blue) resonator with $f_0 = 5.9685$ GHz.}
    \label{measfig}
\end{figure}

\end{document}

%% file: preamble.tex
\usepackage{amsthm}
\usepackage{mathtools}
\usepackage{physics}
\usepackage{xcolor}
\usepackage{graphicx}
\usepackage[left=23mm,right=13mm,top=35mm,columnsep=15pt]{geometry} 
\usepackage[T1]{fontenc}
\usepackage{lipsum}
\usepackage{csquotes}
\usepackage{gensymb}

%% file: abstract.tex

Substrate material imperfections and surface losses are one of the major factors limiting superconducting quantum circuitry from reaching the scale and complexity required to build a practicable quantum computer. One potential path towards higher coherence of superconducting quantum devices is to explore new substrate materials with a reduced density of imperfections due to inherently different surface chemistries. Here, we examine two ternary metal oxide materials, spinel (MgAl$_2$O$_4$) and lanthanum aluminate (LaAlO$_3$), with a focus on surface and interface characterization and preparation. Devices fabricated on LaAlO$_3$ have quality factors three times higher than earlier devices, which we attribute to a reduction in interfacial disorder. MgAl$_2$O$_4$ is a new material in the realm of superconducting quantum devices and, even in the presence of significant surface disorder, consistently outperforms LaAlO$_3$. Our results highlight the importance of materials exploration, substrate preparation, and characterization to identify materials suitable for high-performance superconducting quantum circuitry.

%% file: intro.tex
\section{Introduction} \label{intro}

Silicon and sapphire are the current workhorse substrates for superconducting quantum devices.\cite{deleon2021, murray2021a} As superconducting devices have advanced, material engineering has played a critical role in increasing resonator quality factor and qubit coherence time, with notable increases due to the removal of disorder at device interfaces.\cite{siddiqi2021, muller2019} In the case of resonators, state-of-the-art devices on silicon and sapphire substrates can have quality factors on the order of $10^6$ in the single photon regime. However, the range of potentially useful (and widely available) substrates has only been explored in a limited fashion.\cite{deleon2021} The focus on new substrate materials is motivated by past success in substrate engineering and more recent work on utilizing less common superconducting materials such as tantalum.\cite{place2020} Exploring new materials has the potential to shed new light on the microscopic origins of decoherence, ultimately leading to improved device function.\cite{deleon2021} Currently, material and interfacial losses are the most prominent factors hindering the development of a useful quantum computer.\cite{murray2021a, altoe2020}


Alternative substrate materials for superconducting devices have been most investigated in relation to hybrid quantum devices, where superconducting elements are coupled to optical, mechanical, or spin degrees of freedom. Superconducting circuitry has been fabricated on substrates such as amorphous SiN$_x$ \cite{paik2010}, Y$_2$SiO$_5$ \cite{dold2019}, GaAs \cite{kopke2014, tournet2016}, GaN \cite{kervinen2018}, SiGe \cite{sandberg2021}, and diamond \cite{lee2019} with the general trend being equivalent or lower measures of resonator quality factor (or qubit coherence times) compared to state-of-the-art devices grown on silicon or sapphire. 


In the case of GaAs, quality factors of around 40,000 at $\langle n_{ph} \rangle = 100$ (where $\langle n_{ph} \rangle$ is the average number of photons in the resonator) have been linked to the intrinsic piezoelectric response leading to microwave losses.\cite{scigliuzzo2020} Away from substrates geared towards hybrid quantum devices, there are isolated reports of superconducting resonators fabricated on La$_{0.3}$Sr$_{0.7}$Al$_{0.65}$Ta$_{0.35}$O$_3$ (LSAT), LaAlO$_3$ \cite{arzeo2014, benz1998}, MgO \cite{arzeo2014}, and more recently van der Waals materials such as h-BN \cite{wang2021} have entered the picture. However, the literature on alternative substrates remains sparse compared to reports on silicon or sapphire substrates or skews towards high-T$_\text{c}$ devices. These reports contain few mentions of surface preparation, suggesting interfacial losses may play a significant role in the limited device performance. Additionally, devices may not be tested in the low power (single photon) regime where superconducting quantum devices operate.

Lanthanum aluminate (LaAlO$_3$) is a widely explored material as a substrate for the growth of high-T$_\text{c}$ superconducting devices owing to the small lattice mismatch with YBCO (yttrium barium copper oxide) and low losses at millikelvin temperatures and microwave frequencies.\cite{zhuravel2010} More recently, LaAlO$_3$ has found broad use in all-oxide electronics and as a substrate for perovskite heterostructures.\cite{biswas2017,boschker2017,braun2020a} In particular, LaAlO$_3$ films on SrTiO$_3$ are of interest due to the presence of superconductivity and a tunable 2D electron gas at the interface between two band insulators.\cite{ohtomo2004} Another promising substrate material is spinel (MgAl$_2$O$_4$), which has attracted wide interest as a substrate for the growth of III-V nitrides and the post-transition metal oxide ZnO.\cite{chen2000} More generally, a wide range of magnetic materials adopt spinel or inverse spinel lattices making MgAl$_2$O$_4$ a popular substrate choice when growing materials such as Co$_3$O$_4$.\cite{vaz2009, suzuki} Importantly, recent years have seen significant advances in surface preparation of ternary metal oxides leading to sharper interfaces, higher-quality devices, and broader functionalities.\cite{boschker2017,biswas2017,braun2020a}

We characterize, fabricate and operationally test aluminum superconducting coplanar waveguide resonators on LaAlO$_3$(100) and MgAl$_2$O$_4$(100) substrates. For LaAlO$_3$, we show that adopting surface preparation protocols developed for all-oxide electronics leads to substantially increased quality factors compared to earlier works and we highlight pathways for further enhancement. For MgAl$_2$O$_4$ substrates, atomic force microscopy (AFM) shows only the faintest signs of surface ordering after \textit{ex-situ} preparation, but the aluminum film quality and low-temperature device performance exceeds LaAlO$_3$. Our initial tests of MgAl$_2$O$_4$ hint at a promising material that may find broader use in superconducting devices.

%% file: experiment.tex
\section{Experimental}\label{experimental}

LaAlO$_3$(100) ($25.4\times25.4\times0.5$ mm) and MgAl$_2$O$_4$(100) ($20.0\times20.0\times1.0$ mm) substrates (MTI Corp.) were rinsed with VLSI grade acetone, isopropanol, and Milli-Q deionized (DI) water. Quartz glassware was used for all wet chemical steps and was pre-cleaned by isopropanol ultrasonication followed by DI rinsing. LaAlO$_3$ and MgAl$_2$O$_4$ samples were loaded into an alumina boat and placed in a tube furnace (Nabertherm C 530). Samples were annealed under flowing oxygen (0.5 l/min) at 1473 K for 3 hours (LaAlO$_3$) and 1373 K for 5 hours (MgAl$_2$O$_4$), with a heating/cooling rate of 3.3 K/min. LaAlO$_3$ samples underwent a slight color change after high-temperature annealing, turning from clear to a faint sepia tone while remaining transparent. After annealing, LaAlO$_3$ was immersed in DI water and ultrasonicated for 120 min to produce a more uniform surface termination as described by Kim et al.\cite{kim2019}
 
Atomic force microscopy (AFM) was performed using Asylum Instruments Cypher and Bruker Dimension XR instruments. The Bruker AFM was operated in PeakForce tapping mode using silicon nitride ScanAsyst-Air AFM tips ($f$ = 70 kHz). On the Cypher AFM, Tap300GD-G model AFM probes ($f$ = 300 kHz) were used. Amplitude, height, and phase channels were recorded. 

 
X-ray photoelectron spectroscopy (XPS) measurements were obtained using a monochromatic Al source (1486.6 eV, Kratos Axis Ultra). As both substrates are bulk insulators, an electron flood gun was used for charge neutralization and the C 1s core level was used as an internal reference. Smaller $5\times5$ mm$^2$ samples were used for XPS measurements to avoid electron-induced defect formation. Neither sample was subjected to heating or ion bombardment before XPS measurement.

Aluminum films were simultaneously deposited on both substrates at room temperature in a Plassys MEB 550 S electron beam evaporator. Film thickness was monitored by a quartz crystal microbalance (QCM) at a rate of 0.5 nm/s; target thickness of 100 nm. X-ray reflectivity (XRR) measurements were performed using a Rigaku SmartLab (Cu anode) with incident angles between 0$\degree$ and 2$\degree$; data was analyzed using Motofit.\cite{nelson2006} As the approximate thickness is known from the QCM, fit parameters were allowed to vary within 10$\%$ of accepted values (thickness, densities).

Films were patterned by spin coating 1.5 $\mu$m AZ1512 HS positive photoresist, followed by direct-write lithography on a Heidelberg $\mu$PG101, developed for 45 seconds using AZ726 MIF followed by etching in a solution containing 21$\%$ DI water, 73$\%$ phosphoric acid, 3$\%$ acetic acid, and 3$\%$ nitric acid (v/v) to selectively etch aluminum. The remaining resist was lifted off in hot acetone. Input impedance and resonance frequencies were calculated using a transmission-line coupled CPW resonator model developed by Besedin et al.\cite{besedin2018} Each chip contains five resonators from 4 to 8 GHz. Low-temperature testing was carried out in a Bluefors dilution fridge, with the mixing chamber at T = 30 mK during measurement.  To reach the single-photon regime (approximately -140 dBm input power), the input signal was attenuated by a total of -120 dBm of hardware attenuation (-60 dBm inside and outside the fridge). A vector network analyzer (Agilent Technologies N5232A PNA-L) was used to generate microwave signals from -30 dBm to 16 dBm. 



%% file: results.tex
\section{Results}\label{results}

Figure 1a shows monochromatic XPS spectra obtained on a MgAl$_2$O$_4$(100) substrate before and after tube furnace annealing. High-resolution XPS measurements show single component Al 2p and Mg 2p peaks while the O 1s peak contains a high binding energy shoulder due to hydroxyls (Figure 1b). The most prominent change after annealing is a small increase in Mg, Al, O peak intensity and reduction in the residual carbon coverage and surface hydroxyls. Neglecting carbon, we determine the composition after annealing to be 15$\%$ Mg, 31$\%$ Al, and 54$\%$ O which is close to the expectation for bulk MgAl$_2$O$_4$. 

XPS measurements of the LaAlO$_3$(100) surface at various stages of surface treatment are shown in Figure 2. Survey scans confirm La, Al, and O peaks with adventitious carbon the sole outlier. Compared to MgAl$_2$O$_4$(100), we observe intensity changes and chemical shifts at each stage of the process. For the untreated and annealed preparations, the O 1s peak contains a prominent high binding energy shoulder consistent with surface or adsorbed hydroxyls. Compared to the untreated sample, tube furnace annealing results in a shift of the La 3d, Al 2p, and O 1s levels to lower binding energies ($<$ 0.23 eV shift). After DI leaching, the core level shifts are reversed, with the La 3d, Al 2p, and O 1s levels located at slightly greater binding energies than the untreated surface ($<$ 0.34 eV shift). Furthermore, after DI leaching, the O 1s level contains a chemically distinct species at 532.1 eV. Given the final processing step occurs in water, we attribute this O 1s species to residual adsorbed water or a chemically distinct hydroxyl species. Beyond water adsorption, the chemical composition within the sampling depth shifts with processing. After annealing the composition of the probed depth (ignoring carbon) is 17$\%$ La, 21$\%$ Al, and 62$\%$ O compared to an expected stoichiometric composition of 20$\%$ La, 20$\%$ Al, and 60$\%$ O. After the DI leaching, we observe a net loss of La with the composition within the probed depth of approximately 14$\%$ La, 22$\%$ Al, and 64$\%$ O. This change is indicative of the removal of surface LaO, as intended, and this is explicitly shown in Figures 3a-d. Therefore, after DI leaching the surface has a nominal AlO$_2$ surface termination. We note that the change in surface composition likely reflects a more complex structure given the polar nature of the (100) surface. 

Tapping mode AFM imaging of annealed LaAlO$_3$(100) shows that the surface is flat with a mix of terrace shapes and sizes (Figure 3a). Non-uniform steps heights between 0.3 nm and 0.5 nm indicate a mixed surface termination, a common occurrence on annealed perovskite oxides.\cite{biswas2017} Turning to the phase-contrast image shown in Figure 3b, we observe distinct regions of contrast within the same terrace, confirming the mixed surface termination picture determined by topographical measures. After DI leaching, we observe a more uniform surface termination for LaAlO$_3$ which we attribute to the removal of LaO (Figure 3c). On the DI leached surface we observe regular step heights of 0.36 nm and more uniform phase contrast (Figure 3d). Comparing AFM phase images taken before and after DI leaching, we observe a 16$\%$ decrease in the prevalence of the LaO at the surface after leaching. This correlates well with the observed reduction in La as seen by XPS and provides further evidence that the surface is nominally AlO$_2$ terminated.

For the MgAl$_2$O$_4$(100), the as-received substrates show no signs of surface order and have an RMS roughness on the order of 17 nm (not shown). After high-temperature annealing, only very faint signatures of structural order are observed in AFM images such as a weak cross-hatched structure near a surface impurity (Figure 3e,f). While images of MgAl$_2$O$_4$(100) were significantly more difficult to obtain compared to LaAlO$_3$(100) due to surface roughness, annealing substantially reduces the RMS roughness to around 5 nm.

While the aluminum film thickness is roughly known from QCM measurements, this estimate does not provide information on the buried interface or deposited film. Figure 4 shows AFM images and x-ray reflectivity (XRR) measurements of aluminum films grown on both substrates. AFM topography of the aluminum film (with native oxide) shows that the RMS roughness is higher for films on LaAlO$_3$ (3.2 nm) compared to MgAl$_2$O$_4$ (2.1 nm). Phase images (Figure 4b, d) clearly resolve aluminum grains on LaAlO$_3$, whereas phase-contrast on MgAl$_2$O$_4$ is much more reduced, indicating the grain structure is more compact. From a simple XRR film-substrate model, we estimate a spatially averaged film thickness of 106 nm for both films. While the film thickness is consistent between XRR measurements, we do observe differences in the decay and amplitude of oscillations in the XRR data. In particular, the Kiessig fringes are more clearly resolved for aluminum films on MgAl$_2$O$_4$ compared to LaAlO$_3$ indicating that the buried Al/LaAlO$_3$ interface is rougher leading to an increase in diffuse scattering. 

Figure 5 shows the schematic layout of both the MgAl$_2$O$_4$ and LaAlO$_3$ resonator circuits and optical micrographs of the final fabricated devices. To match the transmission line impedance with the 50 $\Omega$ input line, gap (S) and width (W) dimensions of S = 6 $\mu$m and W = 18 $\mu$m (S = 12 $\mu$m, W = 6 $\mu$m) were used for MgAl$_2$O$_4$ (LaAlO$_3$). These dimensions were chosen to keep the total transmission line width (2S + W) constant at 30 $\mu$m across both devices. This allows for a more accurate operational comparison of the two devices, as the scale of features impacts device performance.\cite{mcrae2020} Optical micrographs of the wet-etched structures are shown in Figure 5b, d. In the case of LaAlO$_3$(100), the optical image shows slight contrast variations which can be attributed to the twinning of the substrate due to distortion from cubic to rhombohedral symmetry below 817 K.\cite{bueble1998a}

Figure 6 shows the microwave transmission responses, which were measured by sweeping the input microwave frequency across the resonance frequency of each resonator and recording the complex transmission, S$_{21}$, through the resonator. The chips were mounted into copper boxes and cooled to 30 mK for measurements at varying input powers corresponding to cavity photon numbers ($n_{ph}$) between 0.1 and 10,000. From the device geometries and measured complex transmission at the resonance frequencies, we determine the relative dielectric permittivities of each substrate (between 4 - 8 GHz) to be $\epsilon_r = 24.32$ for LaAlO$_3$ and $\epsilon_r = 8.25$ for MgAl$_2$O$_4$ indicating no strong shift upon cooling.

At large powers, corresponding to $n_{ph} = 10,000$, we observe strong resonances for each chip and individual resonators. In the case of LaAlO$_3$(100) substrates, the resonance peak has an asymmetric Fano lineshape, which can be attributed to environmental factors such as cable delay, coupling strength, and impedance mismatch at the transmission line.\cite{probst2015} The amplitude and phase of the S$_{21}$ transmission on resonance are shown in Figure 6a, b. In the single-photon limit, the quality factor of the best LaAlO$_3$ resonator was determined to be $Q_i = 59,000$. Figures 6d, e show the S$_{21}$ transmission characteristic of devices fabricated on MgAl$_2$O$_4$(100). In comparison to LaAlO$_3$ devices, this curve has a more symmetrical Lorentzian shape indicative of near-critical coupling between the microwave feedline and the resonator, and good impedance matching at the feedline. In the single-photon limit, the quality factor of the best resonator on MgAl$_2$O$_4$ is $Q_i = 99,000$. 

The power dependence of the quality factor is shown in Figure 6c, f, where the decrease in quality factor with decreasing input power is assumed to be from the depolarization of two-level systems (TLS). We fit this data by 

\begin{equation}
    \frac{1}{Q_i} = F \delta_\text{TLS} \frac{ \tanh \left( \frac{\hbar \omega_0}{2k_B T} \right) } { (1 + \frac{n_{ph}}{n_c})^\beta } +\delta_\text{other}, 
    \label{TLSmodel}
\end{equation}

where $Q_i$ is the internal quality factor, $F$ is the filling factor, $\delta_\text{TLS}$ is the TLS loss tangent (proportional to the TLS concentration in the resonator), $n_{ph}$ is the average number of photons in the resonator, $n_c$ is the cavity photon number required to saturate a single TLS, $\delta_\text{other}$ accounts for all non-TLS losses, and $\beta$ is a fitting parameter.\cite{burnett2016} From the power-dependent TLS loss model, we extract the product of the filling factor $F$ (of order 1) and the TLS loss tangent ($\delta_{\text{TLS}}$) to be $F\delta_{\text{TLS}}$ = 5.0 × 10$^{-6}$ and $F\delta_{\text{TLS}}$ = 9.1 × 10$^{-6}$ for the best spinel and lanthanum aluminate devices, respectively.

%% file: discussion.tex
\section{Discussion}\label{discussion}

Superconducting circuits are one of the leading quantum computing platforms.\cite{neill2018} Yet, fabrication methods hew towards the optimization of well-known recipes with comparatively little materials innovation. As current device performance is largely limited by material issues, it is imperative to explore broader classes of materials to gauge their suitability for quantum devices.\cite{deleon2021,murray2021a} This approach has recently been highly successful in the case of superconducting materials, where tantalum was shown to dramatically increase coherence times.\cite{place2020} As noted in the introduction, while new substrate materials have been trialled, these efforts have not focused on surface preparation putting these efforts behind established methods for working with silicon and sapphire. Substrate selection and surface preparation are critical steps as they control the material growth mode, grain size (and the influence of boundaries), interfacial strain, and roughness.\cite{campbell1997} Such inhomogeneities may host TLS which reduce quality factors and are well-known sources of decoherence.\cite{martinis2005,muller2019}

Our ex-situ substrate preparation and characterization reveals key chemical and structural details of our devices at different stages of the fabrication process. For LaAlO$_3$, we show that a near-uniform surface termination is achievable using tube furnace annealing and DI water rinsing. XPS measurements confirm the as-prepared surface is La deficient, likely indicating an AlO$_2$ or, more likely, a hydroxylated AlO$_\text{x}$ surface termination. Interestingly, we observe strong core level shifts for this sample depending on the preparation stage. Similar effects have been observed on the related material LaFeO$_3$(100) by Stoerzinger et al. using ambient-pressure XPS.\cite{stoerzinger2017} They attribute the core level shifts to band bending in the presence of adsorbed water, finding that FeO$_2$ layers (in our case AlO$_2$) host a higher coverage of hydroxyls compared to LaO layers. Finally, we note that ultrasonication in DI water increases the coverage of carbon to as-received levels, and may influence aluminum nucleation and film growth. AFM and XRR measurements show that the aluminum film has an RMS roughness of 3.2 nm and weak amplitude Kiessig fringes indicative of buried interface roughness.

Taken together, our results show that devices fabricated on well-ordered LaAlO$_3$(100) surfaces have substantially higher quality factors compared to earlier reports ($Q_i=20,000$ at low powers).\cite{arzeo2014, benz1998} We attribute this increase in quality factor to a reduction of interfacial disorder after adopting treatments developed by the pulsed laser deposition community for oxide heterostructure growth. The explicit role of twin domain boundaries on resonator quality factor remains an open and intriguing question. While the presence of these boundaries may ultimately limit device performance, we find no direct evidence on their role and XRR suggests the buried Al/LaAlO$_3$ interface contains residual disorder. The impact of twin boundaries may be more relevant for complex materials such as YBCO and relatively less critical for elemental superconductors. Therefore, we suggest that a more nuanced view of these twin boundaries must be considered in the future.

Spinel substrates have not previously been explored for use in quantum devices but their initial performance is highly encouraging due to the low value of F$\delta_\text{TLS}$.  In contrast to LaAlO$_3$, XPS of MgAl$_2$O$_4$ shows no pronounced core level shifts, indicating overall simpler surface chemistry, and a clear reduction in the C 1s peak after ex-situ preparation. Intriguingly, AFM measurements of MgAl$_2$O$_4$(100) show only faint signs of surface order after annealing. The tube furnace annealing employed here replicates the initial treatment suggested by Jensen et al.\cite{jensen2015}, however, they further treated MgAl$_2$O$_4$ by annealing in a UHV environment to produce regular terraces hundreds of nanometers wide. While faint, we do note some similarities between the structure in Figure 3f and their samples annealed at intermediate temperatures (1173 K and 1273 K).

Given the limited surface order, it is surprising that AFM and XRR show characteristics of high-quality film growth such as large-amplitude Kiessig fringes and more compact film growth. Considering device figures of merit, those fabricated on MgAl$_2$O$_4$ consistently outperform LaAlO$_3$ across all power regimes. While the films presented here are polycrystalline, single-crystalline Al/MgAl$_2$O$_4$(100) films have been fabricated in the context of microelectronics research.\cite{schweinfest2001a, yu2006a, lin2011a} The high quality of these films is due to the small lattice mismatch (0.25$\%$) between the spinel oxygen sublattice and Al(100) planes which promotes cube-on-cube epitaxy.\cite{lin2011a,schweinfest2001a} Surface preparation under UHV conditions, with the formation of large flat terraces and the removal of adventitious carbon, can be expected to produce devices with higher quality factors by reducing interfacial disorder.

Recent advances in superconducting device fabrication have emphasized the importance of post-processing on final device performance. On silicon substrates, modifications such as trenching and wet etching can lead to dramatically higher quality factor devices.\cite{calusine2018} Buffered oxide etch (BOE) has been used to strongly suppress TLS losses at the exposed substrate-air interface.\cite{altoe2020} As BOE is a widely applied surface preparation technique for LaAlO$_3$ (and related perovskite surfaces) wet chemical post-processing may reduce the impact of TLS on our devices.\cite{biswas2017} Furthermore, we note that after BOE processing, the acid needs to be neutralized by rinsing in deionized water. This water rinse is akin to the water leaching step used for LaAlO$_3$ surface preparation and may facilitate ordering within etched regions. As the native aluminum oxide layer limits etching or attack of the aluminum film this post-processing step is likely to be highly selective at removing surface disorder on the exposed LaAlO$_3$ surface and, potentially, MgAl$_2$O$_4$.

%% file: conclusion.tex
\section{Conclusion}\label{conclusion}

We have fabricated aluminum superconducting resonators on ternary metal oxide substrates LaAlO$_3$(100) and MgAl$_2$O$_4$(100) with a focus on substrate preparation, interface characterization, and testing in the single-photon regime. Even though surface characterization revealed a more highly-ordered surface on LaAlO$_3$(100) substrates, MgAl$_2$O$_4$ devices consistently yielded higher $Q$-factors. Although the quality factors of these resonators are below the state-of-the-art values for the resonators fabricated on Si and sapphire substrates, our results are substantially better compared to earlier reports for LaAlO$_3$ and are the first reported measurements for MgAl$_2$O$_4$. Further research is proposed to identify if these substrates (in particular MgAl$_2$O$_4$) can be used viable alternatives to conventional materials. Our results show that the fabrication, characterization and measurement of quantum devices on alternative substrate materials is key to understanding and minimizing decoherence mechanisms.

%% file: acknowledgements.tex
\section*{Acknowledgements}\label{acknowledgements}

The authors acknowledge the Traditional Owners and their custodianship of the lands on which UQ operates. We pay our respects to their Ancestors and their descendants who continue cultural and spiritual connections to Country. The authors thank Rohit Navarathna for wire bonding. The authors acknowledge the facilities, and the scientific and technical assistance, of the Microscopy Australia Facility at the Centre for Microscopy and Microanalysis, The University of Queensland (Anya Yago, Lachlan Casey, Kevin Jack). This work was performed in part at the Queensland node of the Australian National Fabrication Facility. A company established under the National Collaborative Research Infrastructure Strategy to provide nano and microfabrication facilities for Australia’s researchers.